\documentclass[twocolumn,aps,amsmath,amssymb,prl,superscriptaddress,showpacs,floatfix]{revtex4}%

\usepackage{bm}
\usepackage{epsfig}
\usepackage[usenames]{color}
\usepackage{graphicx}

\definecolor{darkgreen}{rgb}{0,0.6,0}
\definecolor{red}{rgb}{1.0,0.0,0.0}
\definecolor{lightblue}{rgb}{0.93,0.96,1}
\definecolor{darkblue}{rgb}{0.,0.,0.6}

\begin{document}


\title{Viscosity control of the  dynamic self-assembly in  ferromagnetic suspensions}
\author{D.~L.~Piet}
\affiliation{Department of   Engineering  Science and Applied Mathematics
Northwestern University 2145 Sheridan Road, Evanston, Il 60208}

\affiliation{Materials Science Division, Argonne National
Laboratory, 9700 South Cass Avenue, Argonne, IL 60439}

\author{A.~V.~Straube}
\affiliation{Department of Physics, Humboldt University of Berlin, Newtonstr. 15, 12489 Berlin, Germany}
\affiliation{Materials Science Division, Argonne National Laboratory, 9700 South Cass Avenue, Argonne, IL 60439}

\author{A.~Snezhko}
\affiliation{Materials Science Division, Argonne National Laboratory, 9700 South Cass Avenue, Argonne, IL 60439}

\author{I.~S.~Aranson}
\affiliation{Materials Science Division, Argonne National
Laboratory, 9700 South Cass Avenue, Argonne, IL 60439}
\affiliation{Department of   Engineering  Science and Applied Mathematics
Northwestern University 2145 Sheridan Road, Evanston,  Il 60208}


\pacs{87.19.ru, 81.16.Dn, 75.50.Tt}

\begin{abstract}
Recent studies of dynamic self-assembly in ferromagnetic
colloids suspended in liquid-air or liquid-liquid interfaces
revealed a rich variety of dynamic structures ranging from linear
snakes to axisymmetric asters, which exhibit novel morphology of the
magnetic ordering accompanied by large-scale hydrodynamic flows.
Based on controlled experiments and first principle theory, we argue
that the transition from snakes to asters is governed by the
viscosity of the suspending liquid where less viscous liquids favor
snakes and more viscous, asters. By obtaining analytic solutions of
the time-averaged Navier-Stokes equations, we gain insights into the
role of mean hydrodynamic flows and an overall balance of forces
governing the self-assembly. Our results illustrate that the
viscosity can be used to control the outcome of the
dynamic self-assembly in magnetic colloidal suspensions.
\end{abstract}

\date{\today}

\maketitle

Fundamental principles guiding self-assembly in non-equilibrium
colloidal systems continues to attract enormous attention in physics
and engineering communities
\cite{bart,osterman,martin,snezhkoJPCM,fischer,alfons,aubry,glotzer,monica,theor,electro,granick,dobnikar2013}.
The interest is stimulated by the need for creating smart materials
capable of self-assembly,  adaptation, and for the design of tunable
structures that can perform useful tasks at the microscale
\cite{mach1}, including targeted cargo delivery~\cite{delivery1},
stirring in microfluidic devices~\cite{microfluid}, and control of optical properties of the media \cite{optical}.

Studies of dynamic self-assembly in ferromagnetic
colloids dispersed  at liquid-air interfaces~\cite{snezhko2,belkin}
and energized by an alternating (\emph{ac}) magnetic field revealed
highly organized, dynamic linear structures -- magnetic snakes. The
snake emerges spontaneously from a random dispersion of particles in
a certain range of frequencies and amplitudes of the {\it ac}
magnetic field. While for low frequencies of the applied magnetic
field the snakes are immobile, with the increase in frequency they
turn into self-propelled entities~\cite{snezhko4}. Surprisingly,
fundamentally new structures -- localized magnetic asters and arrays
of asters -- emerge when the same colloidal suspension is
confined at the interface between two immiscible liquids and
is energized by the alternating magnetic field~\cite{snezhkoNature}.

Both magnetic snakes and asters generate complex flows in the fluid
and possess magnetic ordering and dynamic organization highly
unfavorable under equilibrium conditions. While magnetic snakes are
essentially linear and comprised of antiferromagnetically ordered
segments of ferromagnetically ordered chains of
microparticles~\cite{snezhko2}, asters develop radial structural
order with the ferromagnetically ordered chains emanating from the
center of each aster~\cite{snezhkoNature}.
The mean flows excited by the snakes and asters have fundamentally
different morphology; snakes create quasi-two dimensional flows with
quadrupole symmetry confined near the surface~\cite{belkin} and asters induce three-dimensional toroidal bulk flows \cite{snezhkoNature}. The main forces that control dynamic
self-assembly in such systems involve not only magnetic
dipole-dipole and steric interactions between the particles but also
nontrivial hydrodynamic forces stemming from deformation of the
interface, viscous drag, and entrainment by the large-scale mean
flow. The striking difference between self-assembled structures in
liquid-air and liquid-liquid systems remained unclear since
both systems were driven similarly. Thus, it is critical to understand the fundamental physical parameters
controlling  the transition between these two distinctive dynamic
states.

In this Letter, we perform a systematic experimental and theoretical
study of the snake-aster transition. It is widely believed that
because the motion of each individual colloidal particle is strongly
overdamped, the viscosity sets only an overall time scale. Moreover,
the motion of fluid is often described by the linear Stokes equation, as,
e.g., in Ref.~\cite{martens-13}, and admits a one-way coupling
between the solvent and the particles, when only the particle
dynamics is influenced by the liquid flow but not vice versa
\cite{straube-11}. For our system, neither of these assumptions is
met, presenting a great challenge for the theory. However, on the basis of
controlled experiments and comprehensive analysis of the first
principle model, we have demonstrated that the viscosity defines the
intricate balance between magnetic forces and hydrodynamic forces
arising from the inertia of the particles and suspending liquid. The
magnitude of these forces is inversely proportional to the
viscosity, which can be independently controlled both in our experiment
and theoretical model. We show that at a given frequency and amplitude of the energizing
\emph{ac} magnetic field, the viscosity of a suspending
liquid controls the transition between snakes and
asters; snakes emerge for smaller viscosities, while asters are favored
in more viscous liquids.

In our experiments with liquid-air systems, the structural
transition from snakes to asters is consistently observed when the
viscosity of suspending liquid is gradually increased. Our
theoretical model, the fully nonlinear Navier-Stokes equations
coupled to the dynamics of individual magnetic particles, is reduced
to a set of closed, time-averaged ordinary differential equations
for  particle positions and orientations interacting via magnetic
forces and effective mean hydrodynamic forces arising due to
oscillation of massive particles in a viscous liquid. The cause of
these mean forces are Stokes drift and Rayleigh streaming. In
contrast to the earlier study based on direct simulation of the
Navier-Stokes equations \cite{belkin3}, here we obtain their
analytic solutions, which provide deep insight into the role of
hydrodynamic flows, their detailed structure, and an overall
balance of forces governing the self-assembly. The model is in good
qualitative (and some times quantitative) agreement with the
experiments.

Our experimental apparatus was similar to that described in
Ref.~\cite{snezhko2}. A ferromagnetic colloidal suspension  was
comprised of nickel microspheres with an average size of
$90\,\mu{\rm m}$ (Alfa Aesar Company). Due to defects in
particles, their magnetic moments are often strongly pinned and the
particles behave as magnetically ``hard'' microspheres. The
particles were dispersed at the liquid-air interface, where they
were supported by a surface tension. To exclude the difference
between the deep and shallow liquid layers, a circular glass beaker
($5 \,{\rm cm}$ in diameter) was filled with  liquid depths of
$5 \,{\rm cm}$ and $5 \,{\rm mm}$.

To vary the viscosity of the liquid, a range of water-sucrose
solutions was prepared \cite{migliori}. The colloidal suspension was
energized by an \emph{ac} magnetic field, $H_{ac}=H_{0}\sin(2\pi f
t)$, with the frequency $f$ and amplitude $H_0 = 200 \, {\rm Oe}$,
applied perpendicular to the interface.

\begin{figure}[t]
\includegraphics[width=8.2cm]{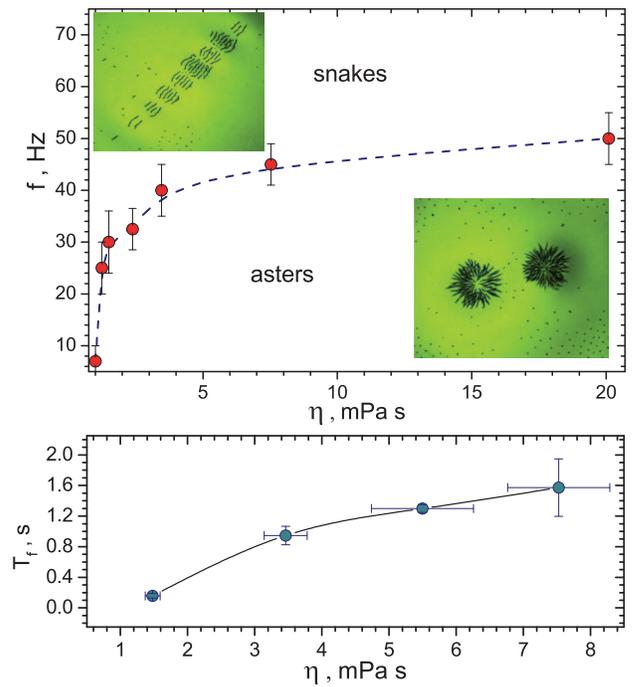}%
\caption{ \emph{Top panel}: Snake-aster phase diagram as a
function of frequency, $f$, and viscosity, $\eta$. The amplitude of
the \emph{ac} magnetic field is $H_0=200\;{\rm Oe}$. \emph{Insets}:
Representative images of a snake (top) and an aster (bottom).
\emph{Bottom panel}: Characteristic time, $T_f$, for the formation of
snakes (or asters) as a function $\eta$ for $f = 40 \;{\rm Hz}$ and
$H_0 =200 \; {\rm Oe}$. For two points the vertical error bars are
smaller than the markers.}
\label{fig1}%
\end{figure}

Selected experimental results are summarized in Fig.~\ref{fig1}. We
observed the formation of magnetic snakes for values of the dynamic
viscosity of the suspending liquid, $\eta$, close to the viscosity
of water, $\eta \approx 1 \, {\rm mPa \, s}$. With a gradual
increase in $\eta$, the snakes give way to asters, as
illustrated in Fig.~\ref{fig1}, top panel. The transition is not
sharp, it is associated with a wide transition region, as indicated
by the error bars. Remarkably, the transition line is almost
parallel to the $\eta$ axis above the viscosity of $\eta \approx 1-2
\, {\rm mPa \, s}$. The bottom panel of Fig.~\ref{fig1} shows a
characteristic time, $T_f$, for the formation of snake or
aster as a function of $\eta$ for $f=40 \; {\rm Hz}$. After
this time, the size of the developed structure almost did not
change, the change of its relative size was within $10\,\%$.
Despite relatively large error bars, $T_f$ gradually increases with
the growth of $\eta$.

To obtain insights into the snake-aster transition, we significantly
extend our model developed in Ref.~\cite{belkin3}. We start with the model
based on the fully nonlinear Navier-Stokes equation in the shallow
water approximation,
\begin{eqnarray}
  \partial_t h + \nabla \cdot \left( h \mathbf{v} \right) &=& 0\,, \label{eq1} \\
  \partial_t \mathbf{v} + \left(\mathbf{v} \cdot \nabla \right) \mathbf{v} &=& \nu \left( \nabla^2 \mathbf{v} - \alpha \mathbf{v} \right)-\nabla h  + \gamma \nabla \nabla^2 h \nonumber \\
  &&+H_0 \sin (\omega t) \sum_i f \left(\mathbf{r} - \mathbf{r}_i \right) \mathbf{P}_i\,, \label{eq2}
\end{eqnarray}
where $\mathbf{v}$ is the two-dimensional (2D), in-plane fluid velocity, $h$ is the surface elevation, $\nu$ is the kinematic viscosity, $\alpha$ is the friction with the bottom of the container, and $\gamma$ is the surface tension.  The last term in Eq.~\eqref{eq2} is representative of forces applied to the surface of the fluid through the particles, where $H_0$ is the amplitude of the \emph{ac} magnetic field, $\omega$ is the frequency, the localized function $f$ defines the shape of the particle, and $\mathbf{P}_i \equiv ( \cos \phi_i, \sin \phi_i )$ is the orientation of the
dipole moment of the $i$th particle. In our study, we neglect the surface tension \cite{footnote} and assume that $f$ is given by delta functions $\delta(\mathbf{r} - \mathbf{r}_i)$, which does not affect the basic physics of self-assembly but, more importantly, makes our model analytically tractable. The variables are scaled as follows: coordinates $\mathbf{r} \to \mathbf{r}/h_0$,
time $t \to t\sqrt{h_0/g}$, velocity $\mathbf{v} \to \mathbf{v} /
\sqrt{gh_0}$, viscosity $\nu \to \nu /h_0\sqrt{g h_0}$, where $g$ is
gravitational acceleration.  In this dimensionless, rescaled equation, $\alpha = h_0 = 1$.

The motion of the particles on the surface of the fluid is described using Newton's equations
    \begin{eqnarray}
        m \ddot{\mathbf{r}}_i + \mu_t \dot{\mathbf{r}}_i &=& \mathbf{F}_i + \mu_t \mathbf{v} - \beta \nabla h\,, \label{eq3} \\
        I \ddot{\boldsymbol{\phi}}_i + \mu_r \dot{\boldsymbol{\phi}}_i &=& \mathbf{T}_i + \kappa H_0 \sin (\omega t) \nabla h \times \mathbf{P}_i\,, \label{eq4}
    \end{eqnarray}
where $m$, $I$, $\mu_t$, $\mu_r$ are the particle mass, moment of inertia, translational and rotational friction coefficients, respectively;
$\beta=mg$, $\mathbf{F}_i=\sum_{j \neq i} \mathbf{F}_{ij}$ and $\mathbf{T}_i=\sum_{j \neq i} \mathbf{T}_{ij}$ are, respectively, the forces and torques on particle $i$ due to magnetic and steric interactions with all other particles, $\mu_t \mathbf{v}$ is the Stokes' drag and $-\beta \nabla h$ is the
movement along the surface gradient from gravity.  The last term in Eq.~\eqref{eq4} is the torque applied to each dipole moment in the direction of the projection of the vertical \emph{ac} field on
deformed surface \cite{belkin3}.

In previous work, Eqs.~(\ref{eq1})-(\ref{eq4}) were solved numerically to model snakes \cite{belkin3}. Here, we first analytically find solutions of
Eqs.~\eqref{eq1} and \eqref{eq2} in an asymptotic limit where we
expand the surface deformation and liquid velocity with respect
to the small parameter $\epsilon$,  $h = h_0 + \epsilon h_1 +
\epsilon^2 h_2+ \mathcal{O}\left( \epsilon^3 \right)$ and
$\mathbf{v} = \epsilon \mathbf{v}_1 + \epsilon^2 \mathbf{v}_2 +
\mathcal{O}\left( \epsilon^3 \right)$. The parameter $\epsilon$ can
be interpreted as the relative deviation of the locus $h({\mathbf r},t)$ of the liquid-air interface
from the equilibrium value $h_0$. Moreover, by using the dimensionless viscosity $\nu$ as
a small parameter, Eqs.~\eqref{eq1} and \eqref{eq2}
were analytically solved up through the first order for the
corresponding surface deformation and 2D velocity fields induced by
each particle individually to yield $h_1({\mathbf r},t) = h_r ({\mathbf r}) e^{\text{i} \omega t} +
\text{c.c.}$ and $\mathbf{v}_1({\mathbf r},t)= \mathbf{v}_r({\mathbf r}) e^{\text{i} \omega t} +
\text{c.c.}$, where $\text{c.c.}$ denotes the complex conjugate. At
the second order, time-averaged solutions $h_2$ and $\mathbf{v}_2$ were sought and a
corresponding analytic expression for $\mathbf{v}_2$, which
determines the mean flow, was obtained \cite{landau,nayfeh}.

Using the explicit solutions $h_1$, $h_2$, $\mathbf{v}_1$, and $\mathbf{v}_2$ of the nonlinear Navier-Stokes
equations, Eqs.~\eqref{eq1} and \eqref{eq2}, we perform the time-averaging of Eqs.~\eqref{eq3} and
\eqref{eq4}. As a result, we arrive at a closed system of ordinary
differential equations for the particles in which all of the details of the
complex hydrodynamic flows are effectively encapsulated in pairwise
interactions
\begin{align}
    m \ddot{\mathbf{r}}_i +  \mu_t \dot{\mathbf{r}}_i = \sum_{j \neq i} \left[ \mathbf{F}_{ij} + \mathbf{s}_{j} + \mu_t \mathbf{v}_2^{(j)} - \beta \nabla h_2^{(j)}\right]\,, \label{eq5}  \\
    I \ddot{\boldsymbol{\phi}}_i + \mu_r \dot{\boldsymbol{\phi}}_i =  \sum_{j\neq i} \left[ \mathbf{T}_{ij} - \frac{\text{i} \kappa H_0}{2} \nabla (h_r^{(j)}-\bar{h}_r^{(j)}) \times \mathbf{P}_i \right]\,. \label{eq6} 
\end{align}
Here, the overline denotes complex conjugate and $\mathbf{s}_{j} = -
2m [\beta\nabla |\nabla h_r^{(j)} |^2 + \mu_t \{ (\mathbf{v}_r^{(j)}
\cdot \nabla) \bar{\mathbf{v}}_r^{(j)} + \text{c.c.} \} ]/(\alpha^2
+ m^2 \omega^2)$ is the Stokes' drift term.  To obtain the Stokes
drift of each particle, we treated each term on the right-hand side
of Eq.~\eqref{eq3} independently. The last term in Eq. \eqref{eq5}
is of much smaller order and can be neglected.

Thus, in contrast to the earlier model \cite{belkin3}, where the dynamics of the particles is determined by Eqs.~(\ref{eq3}) and (\ref{eq4}) coupled to nonlinear equations (\ref{eq1}) and (\ref{eq2}), we suggest a much  simpler and more transparent model in which the particle positions and orientations are described by Eqs.~\eqref{eq5} and \eqref{eq6}. Based on this model, we performed simulations with different numbers of particles ranging from $225$ to $1000$ \cite{Hucht2007}, with an initial configuration on a perturbed square lattice with a uniformly random  orientation of the dipole moment and run on a GPU cluster.
In addition to a significant reduction of computation time, roughly an order of magnitude speed up, the great advantage of our approach is gaining insight into the surface flows as the central ingredient underlying self-assembly.

\begin{figure}[t]
\includegraphics[width=9cm]{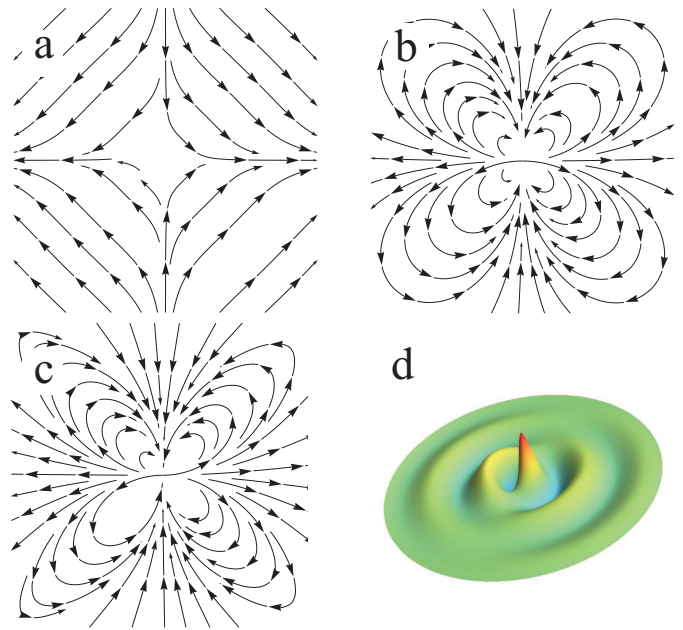}%
\caption{ Mean surface flows ($\mathbf{v}_2$) and first-order surface deformation
($h_r$) induced by a single particle. \emph{Panels (a) and (b):} Quadrupole
streamlines produced by the potential (a) and rotational (b) components of
the mean flow, $\mathbf{v}_2$.  The magnitude of the velocity $\mathbf{v}_2$ decays as
$r^{-3}$. \emph{Panel (c):} Composite of the potential and rotational
flows.    \emph{Panel (d):} First-order surface deformation $h_r$, which decays exponentially with $r$.  The color in the image indicates the height.}
\label{flow_fields}%
\end{figure}

The overall analytic behavior of the mean surface flows, shown in Fig.~\ref{flow_fields}, is similar
to the large-scale quadrupolar flow seen from experiment. These flows are analogous to the
mean flow produced by Rayleigh streaming \cite{riley}. The
first-order flows ($\mathbf{v}_1$) are time dependent, dipolar
flows that oscillate in space and decay out exponentially, ${\mathbf
v}_r({\mathbf r}) \propto \exp(-\text{i}kr)/\sqrt{r}$, with $k\approx \omega - \text{i}\nu
k_1$, $k_1 =(\omega^2+\alpha)/2$; the behavior of $h_r({\mathbf r})$ with $r$ is similar
to that of ${\mathbf v}_r({\mathbf r})$, see Fig.~\ref{flow_fields}(d). The
second-order mean flow, $\mathbf{v}_2$, is time independent and is decomposed into the potential and rotational components, as shown
in Figs.~\ref{flow_fields}(a) and \ref{flow_fields}(b),
respectively. Both these counterparts have a long-ranged quadrupolar
structure with a monotonic power-law decay $\propto r^{-3}$. The
full mean flow is seen in Fig.~\ref{flow_fields}(c).

Note that since the localized shape function was modeled by the delta function, the velocities and surface deformations at all orders
diverge at the center of each particle. These divergences, however,
have no effect on the system because each particle cannot influence
itself and for each pair of particles, a short-range steric
repulsion prevents them from getting close enough to feel the
divergence. Earlier experimental data \cite{belkin} showed that a
mean quadrupole flow was an essential ingredient for the assembly of
snakes. Our model elucidates why this finding is true
for both snakes and asters: Neither of these structures can be
reproduced in simulations via Eqs.~\eqref{eq5} and \eqref{eq6}
unless the Stokes drift and the mean flow (i.e., the fields $h_r$,
${\mathbf v}_r$, and ${\mathbf v}_2$) are properly determined.

\begin{figure}[t]
\includegraphics[width=9cm]{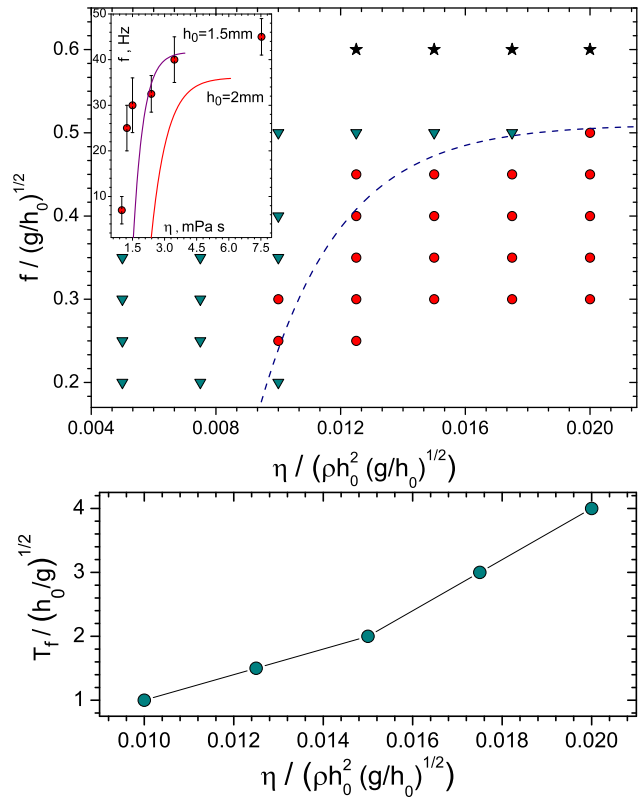}%
\caption{ \emph{Top panel:} Snake-aster phase diagram as a function
of the rescaled frequency and viscosity. Circles,
triangles, and stars are, respectively for asters, snakes, and a
mixture of segments neither forming asters nor snakes. \emph{Inset}:
Comparison of experimental data (circles) and prediction of the
model (solid lines) for $h_0=1.5\;{\rm mm}$ and $h_0=2 \;{\rm mm}$.
\emph{Bottom panel:} Formation time, $T_f$, of a structure as a
function of fluid viscosity for $f(g/h_0)^{-1/2}=0.3$.} 
\label{phase_diagram}%
\end{figure}

A critical test of the model is to recover the crossover  in the
behavior from snakes to asters that was seen experimentally as a
function of the liquid viscosity, $\eta$, and the field
frequency, $f$. The model successfully does so for a range of
values of $\eta$ and $f$, see Fig.~\ref{phase_diagram}, top panel,
where $\alpha = h_0 = \rho = 1$. In qualitative agreement with the
experiment, we observed snakes and asters formed for lower and
higher values of $\eta$, respectively. Moreover, the dependence
of time $T_f$ for the formation of a dynamic structure on
$\eta$ exhibits a trend similar to the experimental one, see Fig.
\ref{phase_diagram}, bottom panel. Note that, in order to avoid
depth dependence, the axes in Fig.~\ref{phase_diagram} remain in
dimensionless quantities. In the case where the viscosity was low
and the frequency was high, the simulations yielded a clumping of
particles primarily due to the lack of friction in the system.
Alternately, when the viscosity was high and the frequency was low,
the particles remained scattered due to overdamping and a lack of
alignment along the changes in the surface height gradient.

\begin{figure}[t]
\includegraphics[width=9cm]{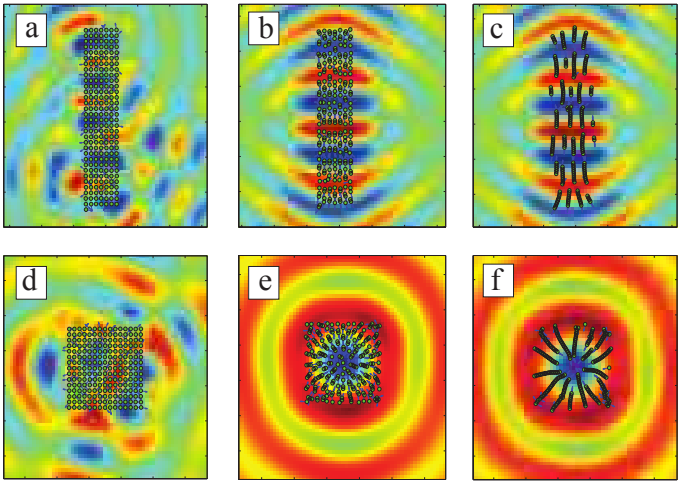}%
\caption{
\emph{Top row:} Snake formation. \emph{Panel (a):} Particles on a
rectangular lattice with random orientation. \emph{Panel (b):}
Magnetic moments align along the surface gradient. Colors represent
surface elevation $h$ (red and blue show maxima and minima, respectively), arrows indicate
particle magnetic moments.  \emph{Panel (c):} Ferromagnetic chains
are formed and are anti-ferromagnetically aligned, creating a snake.
\emph{Bottom row:} Formation of an aster.  \emph{Panel (d):}
Particles on a square lattice with random orientations.  \emph{Panel
(e):} Magnetic moments align along the surface gradient.
\emph{Panel (f):} Ferromagnetically ordered chains are formed and an
aster is assembled.}
\label{formation}%
\end{figure}

Because the mean flow induced by each particle has a long-range nature, it affects the dynamics of all other particles, leading to a highly nontrivial self-organization of the system.
Figure \ref{formation} illustrates the formation of snakes and
asters from an initially disordered distribution of particles.  The
particles were dispersed uniformly inside a rectangle (snakes) or
square (asters) with their magnetic moments oriented randomly.
As Fig.~\ref{formation} shows, asters and snakes are formed after a
short transient ($T_f$), their organization, e.g., anti-ferromagnetic order,
closely resembles the experimental one. Starting from different
initial conditions, e.g., square for the case of snake, often
resulted in the formation of more than one snake or aster.

In conclusion, we have demonstrated that the viscosity of the
suspending liquid strongly affects  the outcome of dynamic
self-assembly and controls the structural transition between
self-assembled structures. Linear snakes are favored for small
viscosities and circular asters for higher viscosities.
Our novel model provides a nontrivial insight into how
the large-scale mean flow -- a nonlinear effect caused by the strong coupling of oscillating particles with the initially equilibrium liquid --  becomes a critical player
that not only determines the shape and organization of the emergent dynamic
structures but also keeps them stable. As we show, both inertia of the liquid and inertia of
the particles are at the heart of the assembly process.

While we were able to reproduce the main observed  phenomenology,
our method also has limitations. One shortfall is that, being a completely two-dimensional model, the shallow
water equations do not capture the liquid jets into the bulk
produced by asters \cite{snezhkoNature}. However, our model is capable of replicating the overall structure of both snakes and asters.
As snakes produce a largely 2D flow, it was natural to expect that they can be recovered from this
model. Asters, however, produce a 3D toroidal flow and their appearance in this model was not expected.
They form due to the propensity of the dipoles to align along the surface
gradient for large characteristic wavelengths, whereas entrainment
by the large-scale toroidal flow is important but less critical. Another reason for the lack of a good quantitative
agreement between the experiment and model (such as in the inset of
Fig.~\ref{phase_diagram}, top panel) is due to the neglected surface tension. Along with
gravity, it presents another mechanism for surface wave generation but
makes the model analytically intractable.

The research was supported by the U.S. DOE, Office of Basic Energy
Sciences, Division of Materials Science and Engineering, under the
Contract No. DE AC02-06CH11357. AVS thanks Argonne's Materials
Theory Institute for support of his visit to Argonne.


\begin{thebibliography}{99}

\bibitem{bart} G.~Whitesides and B.~Grzybowski, Science, \textbf{295}, 2418 (2002).

\bibitem{snezhkoJPCM} A.~Snezhko, J. Phys. Condens. Matter \textbf{23}, 153101 (2011).

\bibitem{martin} J.~E. Martin, Phys. Rev. E {\bf 79}, 011503 (2009); K.~J.~Solis and
J.~E.~Martin, J. Appl. Phys. \textbf{111}, 073507 (2012).

\bibitem{glotzer} S.~C.~Glotzer and  M.~J.~Solomon, Nature Mater. \textbf{6}, 557 (2007).

\bibitem{theor} I.~S.~Aranson and L.~S.~Tsimring, Rev. Mod. Phys. \textbf{78}, 641 (2006);  {\it ibid.} {\it Granular Patterns} (Oxford University Press, Oxford, 2009).

\bibitem{fischer} P.~Tierno, T.~M.~Fischer, T.~Johansen, and F.~Sagues, Phys. Rev. Lett. \textbf{100}, 148304 (2008).

\bibitem{osterman} N.~Osterman \textit{et al.}, Phys. Rev. Lett. \textbf{103}, 228301 (2009); M.~Oettel and S.~Dietrich, Langmuir \textbf{24}, 1425 (2008).

\bibitem{alfons} M.~Leunissen, H.~R.~Vutukuri, and A.~van~Blaaderen, Adv. Mater. \textbf{21}, 3116 (2009).


\bibitem{aubry} N.~Aubry, P.~Singh, M.~Janjua, and S. Nudurupati, Proc. Natl. Acad. Sci. U.S.A. \textbf{105}, 3711 (2008).

\bibitem{monica} G.~Vernizzi and M.~Olvera~de~la~Cruz, Proc. Natl. Acad. Sci. U.S.A. \textbf{104}, 18382 (2007).

\bibitem{electro} M.~V.~Sapozhnikov, Y.~V.~Tolmachev, I.~S.~Aranson, and W.-K.~Kwok, Phys. Rev. Lett. \textbf{90}, 114301 (2003).

\bibitem{granick} J.~Yan, M.~Bloom, S.~C.~Bae, E.~Luijten, and S.~Granick, Nature (London) \textbf{491}, 578 (2012).

\bibitem{dobnikar2013} J.~Dobnikar, A.~Snezhko, A.~Yethiraj, Soft Matter \textbf{9}, 3693 (2013).

\bibitem{mach1}  B.~Grzybowski, \textit{et al.}, Appl. Phys. Lett. \textbf{84}, 1798 (2004).

\bibitem{delivery1} J.~Edd \textit{et al.}, Proceed. of the IEEE/RSJ. Intl. Conf. on Intelligent Robots and Systems {\bf 3}, 2583-2588 (2003).

\bibitem{microfluid} S.~T.~Chang \textit{et al.}, Nature Mater. \textbf{6}, 235 (2007).

\bibitem{optical} S.~K.~Y.~Tang, R.~Derda, A.~D.~Mazzeo, and G.~M.~Whitesides,  Advanced Mater. \textbf{23}, 2413 (2011).

\bibitem{migliori} M.~Migliori, D.~Gabriele, R.~Di~Sanzo, B.~de~Cindio, S.~Correre, J. Chem. Eng. Data \textbf{52}, 1347 (2007).

\bibitem{snezhko2} A.~Snezhko, I.~S.~Aranson, and W.~Kwok, Phys. Rev. Lett. {\bf 96}, 078701 (2006); A.~Snezhko, I.~S.~Aranson, and W.~Kwok, Phys. Rev. E \textbf{73}, 041306 (2006).

\bibitem{belkin} M.~Belkin, A.~Snezhko, I.~S.~Aranson, and W.-K.~Kwok, Phys. Rev. Lett. \textbf{99}, 158301 (2007).

\bibitem{snezhko4}  A.~Snezhko, M.~Belkin, I.~S.~Aranson, and W.-K.~Kwok, Phys. Rev. Lett. {\bf 102}, 118103 (2009).

\bibitem{snezhkoNature} A.~Snezhko, I.~S.~Aranson, Nature Mater. \textbf{10}, 698 (2011).

\bibitem{martens-13} S.~Martens, A.~V.~Straube, G.~Schmid, L.~Schimansky-Geier, and P.~H{\"a}nggi, Phys. Rev. Lett. \textbf{110}, 010601 (2013).

\bibitem{straube-11} A.~V.~Straube, J. Phys. Cond. Matter \textbf{23}, 184122 (2011).


\bibitem{belkin3}  M.~Belkin, A.~Glatz, A.~Snezhko, and I.~S.~Aranson, Phys. Rev. E \textbf{82}, 015301 (2010).

\bibitem{footnote} While lateral capillary forces can influence outcome of colloidal assembly, see, e.g.
R. Di Leonardo, F. Saglimbeni, and G. Ruocco,  \prl {\bf 100},
106103 (2008), in our case capillary interactions (of the order of
pN) are negligible compared to magnetic and hydrodynamic forces (of the order of nN).

\bibitem{landau} L.~D.~Landau and E.~M.~Lifshitz, {\it Fluid Mechanics} (Pergamon Press, Oxford, 1987), 2nd edition.

\bibitem{nayfeh} A.~H.~Nayfeh, Introduction to Perturbation Techniques (Wiley, New York, 1981).


\bibitem{riley} N.~Riley, Theor. Comp. Fluid Dyn. \textbf{10}, 349 (1998).


\bibitem{Hucht2007} A.~Hucht, S.~Buschmann, and P.~Entel, Europhys. Lett. \textbf{77}, 57003 (2007).



\end{thebibliography}
\end{document}